\RequirePackage{lineno}
\documentclass[prd,twocolumn,amsmath,amssymb,showpacs,superscriptaddress,nofootinbib]{revtex4}
\usepackage{graphicx,epsfig}
\usepackage{setspace}
\usepackage{array}
\usepackage{dcolumn} 
\usepackage{bm,color}
\usepackage{multirow}
\usepackage[CJKbookmarks=true,unicode,colorlinks,linkcolor=blue,anchorcolor=blue,citecolor=blue,pdfborder={0 0 0}]{hyperref} 
\newcommand{\BR}{{\cal B}}
\newcommand{\lumi}{{\cal L}}

\newcommand{\jpsi}{J/\psi}

\newcommand{\pip}{\pi^+}
\newcommand{\pim}{\pi^-}
\newcommand{\kz}{K^0}

\newcommand{\ks}{K_S}
\newcommand{\kl}{K_L}
\newcommand{\g}{\gamma}
\newcommand{\ar}{\rightarrow}
\newcommand{\ST}{\rule[-.5em]{0pt}{1.5em}}
\newcommand{\STi}{\rule[-0.5em]{0pt}{1.5em}}
\RequirePackage{xspace}

\usepackage{relsize}
\def\babar{\mbox{\slshape B\kern-0.1em{\smaller A}\kern-0.1em
    B\kern-0.1em{\smaller A\kern-0.2em R}}}
\def\CP                {\ensuremath{C\!P}\xspace}

\begin{document}
\preprint{}

\title{\boldmath Branching fraction measurement of $\jpsi\ar\ks\kl$ and search for $\jpsi\ar\ks\ks$}

\author{\small M.~Ablikim$^{1}$, M.~N.~Achasov$^{9,d}$,
  S. ~Ahmed$^{14}$, M.~Albrecht$^{4}$, M.~Alekseev$^{53A,53C}$,
  A.~Amoroso$^{53A,53C}$, F.~F.~An$^{1}$, Q.~An$^{50,40}$,
  J.~Z.~Bai$^{1}$, Y.~Bai$^{39}$
% ...
  \\
  \vspace{0.2cm}
  (BESIII Collaboration)\\
  \vspace{0.2cm}{\it
    $^{1}$ Institute of High Energy Physics, Beijing 100049, People's Republic of China\\
% ...
$^{4}$ Bochum Ruhr-University, D-44780 Bochum, Germany\\
% ...
$^{9}$ G.I. Budker Institute of Nuclear Physics SB RAS (BINP), Novosibirsk 630090, Russia\\
% ...
$^{14}$ Helmholtz Institute Mainz, Johann-Joachim-Becher-Weg 45, D-55099 Mainz, Germany\\
% ...
$^{39}$ Southeast University, Nanjing 211100, People's Republic of China\\
    $^{40}$ State Key Laboratory of Particle Detection and Electronics, Beijing 100049, Hefei 230026, People's Republic of China\\
% ...
$^{50}$ University of Science and Technology of China, Hefei 230026, People's Republic of China\\
% ...
$^{53}$ (A)University of Turin, I-10125, Turin, Italy; (B)University of Eastern Piedmont, I-15121, Alessandria, Italy; (C)INFN, I-10125, Turin, Italy\\
% ...
$^{d}$ Also at the Novosibirsk State University, Novosibirsk, 630090, Russia\\
% ...
  }\vspace{0.4cm}}

\begin{abstract}
  Using a sample of $1.31\times10^{9}$ $\jpsi$ events collected with
  the BESIII detector at the BEPCII collider, we study the decays of
  $\jpsi \ar \ks\kl$ and $\ks\ks$. The branching fraction of
  $\jpsi\ar\ks\kl$ is determined to be
  $\BR(\jpsi\ar\ks\kl)=(1.93\pm0.01~(\rm
  {stat.})\pm0.05~(\rm{syst.}))\times10^{-4}$, which significantly
  improves on previous measurements. No clear signal is observed for
  the $\jpsi\ar\ks\ks$ process, and the upper limit at the 95\%
  confidence level for its branching fraction is determined to be
  $\BR(\jpsi\ar\ks\ks)<1.4\times10^{-8}$, which improves on the
  previous searches by 2 orders in magnitude and reaches the order
  of the Einstein-Podolsky-Rosen expectation.
\end{abstract}

\pacs{13.66.Bc, 13.25.Gv, 03.65.Vf}

\maketitle

\frenchspacing
\section{Introduction}\label{Introduction}

The charmonium state $\jpsi$ with a mass below the open charm
threshold decays to light hadrons through the annihilation of
$c\bar{c}$ into one virtual photon, three gluons or one photon and two
gluons. The $\jpsi$ decaying to $K_SK_L$ proceeds via the first two
processes, thereby providing valuable information to understand the
nature of $\jpsi$ decays. The available measurements of its branching
fraction, $\BR(\jpsi\ar\ks\kl)$, based on 57.7 million $\jpsi$ events
collected at BESII~\cite{ksklbes2} and 24.5 million $\psi(3686)$
events at CLEO~\cite{kkcleo}, are given by
$(1.82\pm0.04\pm0.13)\times10^{-4}$ and
$(2.62\pm0.15\pm0.14)\times10^{-4}$ respectively. Due to the
discrepancy between these two measurements, the world average value in
the particle data group (PDG)~\cite{PDG2016} has quoted a relative
precision of 19\%, which limits the precise understanding of $\jpsi$
decay mechanisms.

In the \CP-violating decay of $\jpsi$ to $\ks\ks$, the two identical
bosons from the decay would need to form an antisymmetric state, and
the process would be ruled out according to Bose-Einstein statistics.
However, according to the Einstein-Podolsky-Rosen (EPR)~\cite{epr}
paradox, the quantum state of a two-particle system cannot always be
decomposed into the joint state of the two particles. Thus the
spacelike separated coherent quantum system may also yield a sizable
decay branching fraction of $\jpsi\ar\ks\ks$ at the $10^{-8}$
level~\cite{roos}. In this way, the $\ks\ks$ system can be used to
test the EPR paradox versus quantum theory. There also might be a
small possibility to have a $\ks\ks$ final state due to \CP
violation. In the $\kz\hbox{--}\bar{\kz}$ oscillation model~\cite{lihb}, the
\CP violating branching fraction of $\jpsi\ar\ks\ks$ is calculated to
be $(1.94\pm0.20)\times 10^{-9}$. The MARKIII experiment searched for
the decay $\jpsi\ar\ks\ks$ with 2.7 million events, and the upper
limit was determined to be $\BR(\jpsi\ar\ks\ks)<5.2\times 10^{-6}$ at
the 90\% confidence level~(C.L.)~\cite{kkmark}.  Based on 57.7 million
$\jpsi$ events collected at the BESII detector, the upper limit on the
branching fraction was improved to be $1.0\times 10^{-6}$ at the 95\%
C.L.~\cite{ksksbes2}, which is still far from the expectations from
EPR and $\kz\hbox{--}\bar{\kz}$ oscillation.

The world's largest $\jpsi$ sample with $1.31\times 10^{9}$ events was
accumulated at BESIII during 2009 and 2012~\cite{jpsino}. In this
paper, we measure the branching fraction of $\jpsi\ar\ks\kl$, and also
search for the \CP violating decay $\jpsi\ar\ks\ks$.

\section{Apparatus and Monte Carlo simulation} \label{Detector}

The Beijing Spectrometer III~(BESIII), located at the double-ring
$e^+e^-$ Beijing Electron Positron Collider (BEPCII), is a general
purpose detector as described in Ref.~\cite{Ablikim:2009vd}. It covers
93\% of $4\pi$ in geometrical acceptance and consists of four main
detectors. A 43-layer small-cell, helium gas based drift chamber,
operating in a 1.0 (0.9)~T solenoidal magnetic field in 2009 (2012),
provides an average single-hit resolution of 135~$\mu$m. A
time-of-flight system, composed of 5 cm thick plastic scintillators
with 176 bars of 2.4 m length, arranged in two layers in the barrel
and 96 fan-shaped counters in the end caps, has a time resolution of
80 ps~(100 ps) in the barrel (end caps) region providing 2$\sigma$
$K/\pi$ separation for momenta up to 1.0 GeV/$c$. An electromagnetic
calorimeter, which consists of 5280 CsI(Tl) crystals arranged in a
cylindrical structure in the barrel and 480 crystals in each of the
two end caps, provides an energy resolution for a 1.0 GeV/$c$ photon
of 2.5\% in the barrel region and 5\% in the end caps. The position
resolution is 6 mm (9 mm) in the barrel (end caps). A muon counter
system, which consists of resistive plate chambers arranged in nine
barrel and eight end-cap layers, provides 2.0 cm position resolution.

The optimization of event selection criteria, the determination of
detection efficiencies, and the estimation of background are performed
by means of Monte Carlo (MC) simulations. The KKMC~\cite{kkmc}
generator is used to simulate the $J/\psi\rightarrow \kz\bar{\kz}$
process. The angular distribution of the $\kz$ or $\bar{\kz}$ is
generated to be proportional to $\sin^{2}\theta$, where $\theta$ is
the polar angle in the laboratory system.  In the MC simulation, the
interference between the $\jpsi$ resonance decay and the continuum
process is ignored. A GEANT4-based~\cite{Agostinelli:2003hh,
  Allison:2006ve} detector simulation software, which includes the
geometric and material description of the BESIII spectrometer, and the
detector response, is used to generate the MC samples.  The background
is studied with a MC sample of $1.23\times10^{9}$ inclusive $\jpsi$
decays, in which the known decays are generated with the
EvtGen~\cite{evtgen1, evtgen2} generator by setting the branching
fraction to the values in the PDG~\cite{PDG2016} and the remaining
unknown decays are generated with the LUNDCHARM~\cite{Chen:2000}.

\section{\boldmath Branching fraction measurement of $\jpsi\ar\ks\kl$}~\label{kskl}

\begin{figure}[htbp]\centering
\includegraphics[width=0.45\textwidth]{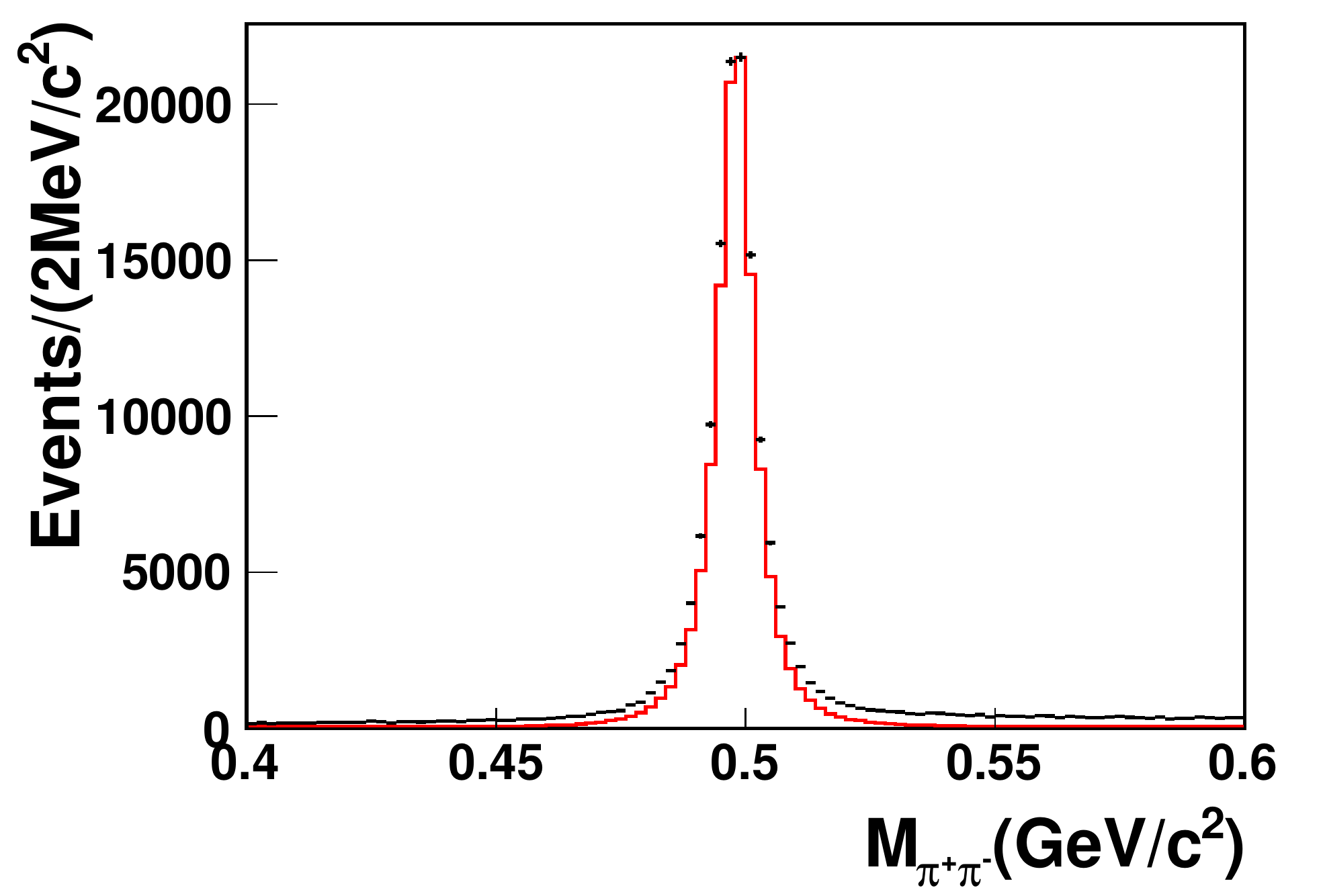}
\caption{The distribution of $M_{\pip\pim}$. The (black) crosses are
  from data, and the (red) histogram represents the signal MC sample.}
\label{mpipi}
\end{figure}

The $\ks$ candidate is reconstructed from its charged $\pip\pim$ final
state, while the $\kl$ is assumed not to decay in the detector leaving
only the signature of missing energy.  The $\ks$
candidates are reconstructed with vertex-constrained fits to pairs of
oppositely charged tracks, assumed to be pions, whose polar angles
satisfy the condition $|\cos\theta|<0.93$.  Only one $\ks$ candidate
is accepted in each event. The $\ks$ candidates are required to
satisfy $L>$1 cm and $L/\sigma_{L}>2$, where $L$ is the distance
between the common vertex of the $\pip\pim$ pair and the interaction
point and $\sigma_{L}$ is its uncertainty. The invariant mass of the
$\pip\pim$ pair, $M_{\pip\pim}$, shown in Fig.~\ref{mpipi}, is required
to satisfy $|M_{\pip\pim}-M_{\ks}|<18$ MeV/$c^{2}$, where $M_{K_S}$ is
the $K_S$ nominal mass~\cite{PDG2016}. There should be no extra tracks
satisfying $|\cos\theta|<0.93$, within 1 cm of the interaction point
in the transverse direction to the beam line and 10 cm of the
interaction point along the beam axis.  In order to suppress $\gamma$
conversion background, the angle between the two charged tracks,
$\theta_\text{ch}$, is required to satisfy $\theta_\text{ch}>15^{\circ}$.

The same event selection criteria are applied to the inclusive MC
sample. The major potential backgrounds are $\jpsi\ar\pi^{0}\ks\kl$
and $\jpsi\ar\g\ks\ks$ events, but the leakage of their $K_S$
momentum~($P_{\ks}$) spectra into the signal region is smooth and
tiny.

\begin{figure}[htbp]\centering
\includegraphics[width=0.45\textwidth]{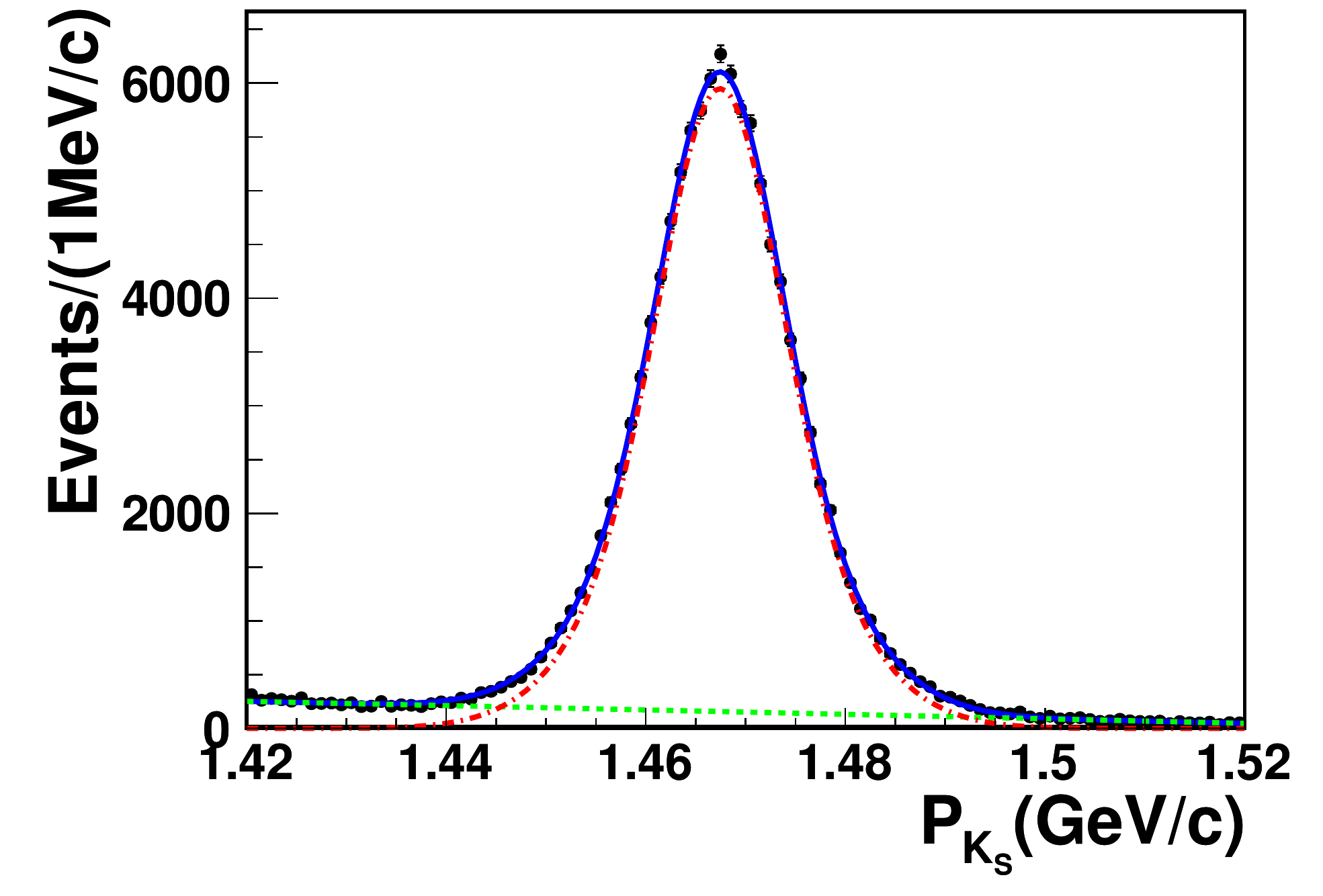}
\caption{The momentum distribution of $\ks$ in the $e^+e^-$ rest
  frame. The (black) crosses are from data, and the (blue) solid line
  is the fit result. The (red) dash-dotted line is the signal, and the
  (green) dashed line is background.}
\label{pksfit}
\end{figure}

The $\jpsi\ar\ks\kl$ signal yield is determined from a
maximum likelihood fit to the $P_{\ks}$ distribution, as shown in
Fig.~\ref{pksfit}. In the fit, the signal shape is described by a
double Gaussian function with a common mean value and two different
widths. The background shape is represented by a second-order
Chebychev polynomial function.

\begin{figure}[htbp]\centering
\includegraphics[width=0.45\textwidth]{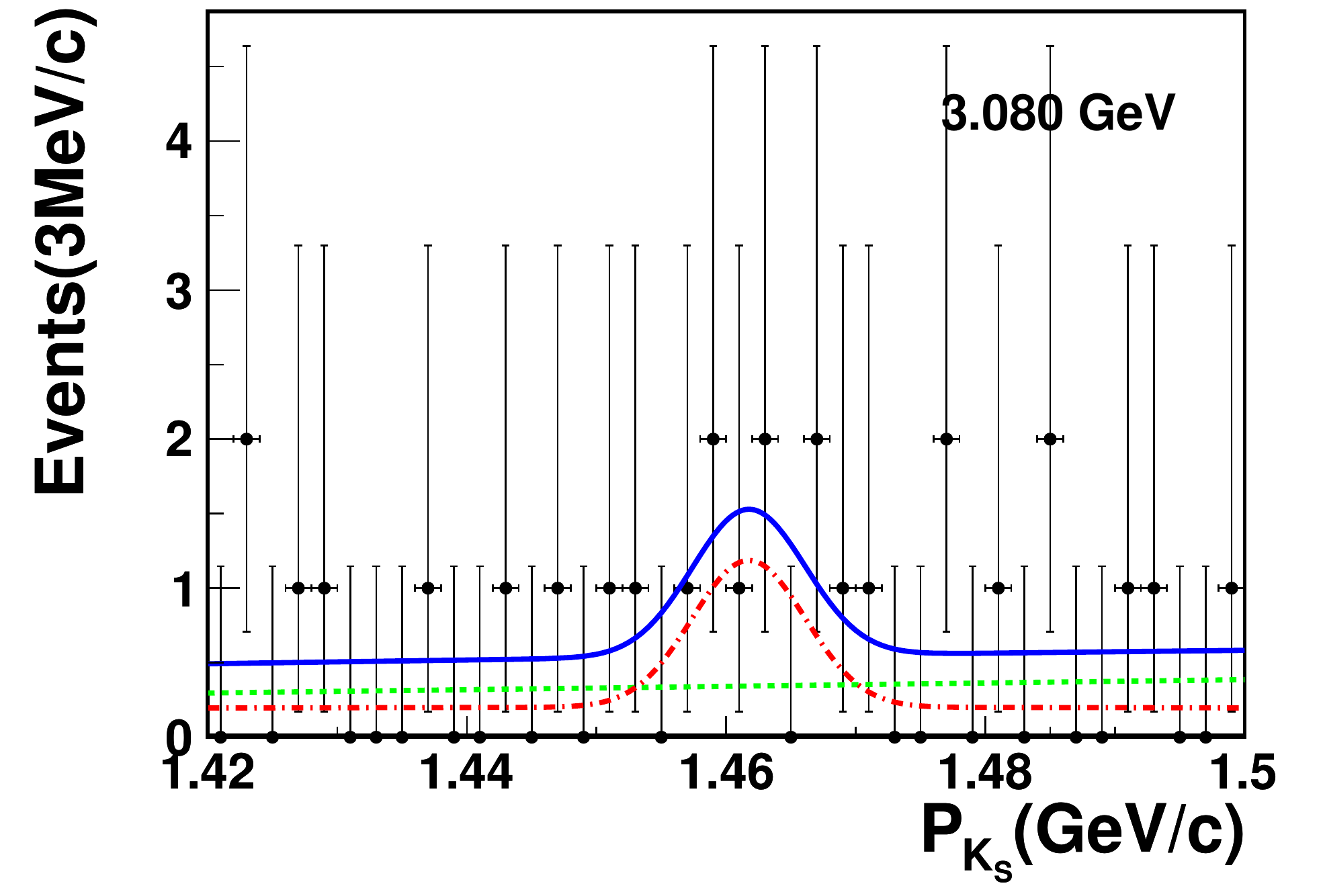}
\caption{The $\ks$ momentum distribution for data taken at
  $\sqrt{s}= 3.080$ GeV. The (black) crosses are data, and the (blue)
  solid line is the fitting result. The (red) dash-dotted line
  corresponds to the signal, and the (green) dashed line represents
  the background.}
\label{ksklcont}
\end{figure}

The continuum process $e^+e^-\ar\ks\kl$ is studied with a data set of
30.0~pb$^{-1}$ taken at 3.080 GeV.  The same selection criteria are
applied. The result of the maximum likelihood fit to the $P_{\ks}$
distribution is shown in Fig.~\ref{ksklcont}. In the fit, the signal
function is the same as that used in the fit of $\jpsi$ data.  The
background shape is represented by a first-order Chebychev polynomial
function.

The event selection efficiencies are assumed to be the same at 3.080
GeV and the $\jpsi$ resonance. The continuum contribution to the
$\jpsi$ resonance region is estimated from
\begin{equation}
N_{\rm {cont}}^{\jpsi}=N_{\rm{obs}}^{3.080}\cdot\frac{{\cal L}\cdot s'^3}{{\cal L}'\cdot s^3},
\end{equation}
where $N_{\rm{obs}}^{3.080}$ is the signal yield at 3.080~GeV,
${\cal L}$ and ${\cal L}'$ are the luminosities collected at the
$\jpsi$ and at 3.080~GeV, determined with $e^+e^-\ar\gamma\gamma$
events~\cite{jpsino}, while $s$ and $s'$ correspond to the squares of
center-of-mass energies of $\jpsi$ and 3.080 GeV.  The power law of
the center-of-mass energy follows the $K^+K^-$ cross section slope
measured by \babar~\cite{babarkk}.

Assuming no interference between the $\jpsi$ decay and the continuum process,
the branching fraction is determined from
\begin{equation}
\BR(\jpsi\ar\ks\kl)=\frac{N_{\rm{obs}}^{\jpsi}-N_{\rm{cont}}^{\jpsi}}{\epsilon \cdot N^{\jpsi} \cdot \BR(\ks\ar\pip\pim)},
\end{equation}
where $N_{\rm{obs}}^{\jpsi}$ is the number of signal events obtained
in the $\jpsi$ sample, $\epsilon$ is the event selection efficiency,
$N^{\jpsi}$ is the number of $\jpsi$ events~\cite{jpsino} and
$\BR(\ks\ar\pip\pim)$ is the branching fraction of
$\ks\ar\pip\pim$. Table~\ref{ksklresult} summarizes the values used in
the calculation, and $\BR(\jpsi\ar\ks\kl)$ is determined to be
$(1.93\pm0.01)\times 10^{-4}$, where the quoted uncertainty is purely
statistical.

\begin{table}[htpb]
    \centering
    \caption{Numbers used in the branching fraction calculation for the $\ks\kl$ channel, where the uncertainties are
    statistical only.}
        \begin{tabular}{p{3cm}<{\centering}p{2.75cm}<{\centering}p{2cm}<{\centering}}
        \hline
                                                     &3.097~GeV~($\jpsi$)              &3.080 GeV                       \ST  \\
        \hline
        $N_{\rm{obs}}$                               &$110203 \pm 504$                  &$13\pm5$                          \\
        $\epsilon~(\%)$                              &62.9                            &62.9                              \\
        $\lumi~(\rm{pb}^{-1})$                       &394.7                           &30.9                              \\
        $\BR(\ks\ar\pip\pim)$~\cite{PDG2016}         &0.692                           &0.692                             \\
        \hline
        \end{tabular}
    \label{ksklresult}
\end{table}

The systematic uncertainties for the $\BR(\jpsi\ar\ks\kl)$ measurement
include those due to $\ks$ reconstruction, the requirement on
$\theta_\text{ch}$, the fit to the $P_{\ks}$ spectrum, the branching
fraction of the $\ks$ decay, and the number of $\jpsi$ events.

The $\ks$ reconstruction involves the charged track reconstruction of
the $\pip\pim$ pair, the vertex fit and the $\ks$ mass window
requirement. The corresponding systematic uncertainty is estimated
using a control sample of $\jpsi\ar K^{*\pm}(892)K^{\mp}$ events,
where $K^{*\pm}(892)\ar\ks\pi^{\pm}$. The momentum of the $\ks$,
$P_{\ks}$ in $\jpsi\ar\ks\kl$ decay is around 1.46 GeV/$c$; thus only
$\ks$ candidates with momentum larger than 1 GeV/$c$ in the control
sample are considered. The ratio of the reconstruction efficiency of
the data over that in the MC is taken as a correction factor to the
$\ks\kl$ selection efficiency, while the uncertainty of the ratio,
1.4\%, is taken as the systematic uncertainty.

The uncertainty from the $\theta_\text{ch}$ requirement is estimated by varying
the selection range. The range is expanded and contracted by
$5^{\circ}$, and the largest change in the branching fraction with
respect to the nominal value is taken as the systematic uncertainty.

The systematic uncertainty related to the fit method is estimated by
varying the fit range and the background shape simultaneously. The fit
range is expanded and contracted by 8 MeV/$c$.  For the $\jpsi$ data
sample, the background shape is varied from a second-order Chebyshev
polynomial function to a third-order Chebyshev polynomial function and
an exponential function. For the continuum data sample, the background
is replaced by a second-order Chebychev polynomial function. The
largest change in the branching fraction is treated as the systematic
uncertainty.

The branching fraction of $\ks\ar\pip\pim$ is taken from the
PDG~\cite{PDG2016} and its uncertainty is 0.1\%. The number of $\jpsi$
events and its uncertainty are determined with $\jpsi$ inclusive
decays~\cite{jpsino}.

The summary of all individual systematic uncertainties is shown in
Table~\ref{ksklsys}, where the total uncertainty is obtained by adding
the individual contributions in quadrature.

\begin{table}
    \centering
    \caption{Systematic uncertainties for the measurement of branching fraction of the $\ks\kl$ channel.}
        \begin{tabular}{p{3cm}<{\centering}p{3cm}<{\centering}}
        \hline
        Source                  &Uncertainty~(\%)      \ST  \\
        \hline
        $\ks$ reconstruction    &1.4                    \\
        $\theta_\text{ch}$      &1.0                    \\
        Fit to $P_{\ks}$        &1.9                    \\
        $\BR(\ks\ar\pip\pim)$   &0.1                    \\
        $N_{\jpsi}$             &0.6                    \\
        \hline
        \ST Total                   &2.6                    \\
        \hline
        \\
        \end{tabular}
    \label{ksklsys}
\end{table}

\section{\boldmath Search for $\jpsi\ar\ks\ks$}

For $\jpsi\ar\ks\ks$ with $\ks\ar\pip\pim$, the final state is
$\pip\pim\pip\pim$.  The candidate events are required to have at
least four charged tracks whose polar angles satisfy
$|\cos\theta|<0.93$. The $\ks$ candidates are reconstructed by secondary
vertex fits to all oppositely charged track pairs assuming them to
be pions, and the $\pip\pim$ invariant mass must be within 18
MeV/$c^2$ from the $\ks$ nominal mass. The $\ks$ candidates must have
a momentum within the range of [1.40,\ 1.60] GeV/$c$. In order to suppress the
non-$\ks$ backgrounds, the decay length over its
uncertainty ($L/\sigma_{L}$) has to be larger than 2.0. Each event must
have at least two $\ks$ candidates. If there are more than
two $\ks$ candidates, the combination with the
smallest sum of $\chi^{2}$ of the secondary vertex fits is selected.

\begin{figure}[htbp]\centering
\includegraphics[width=0.45\textwidth]{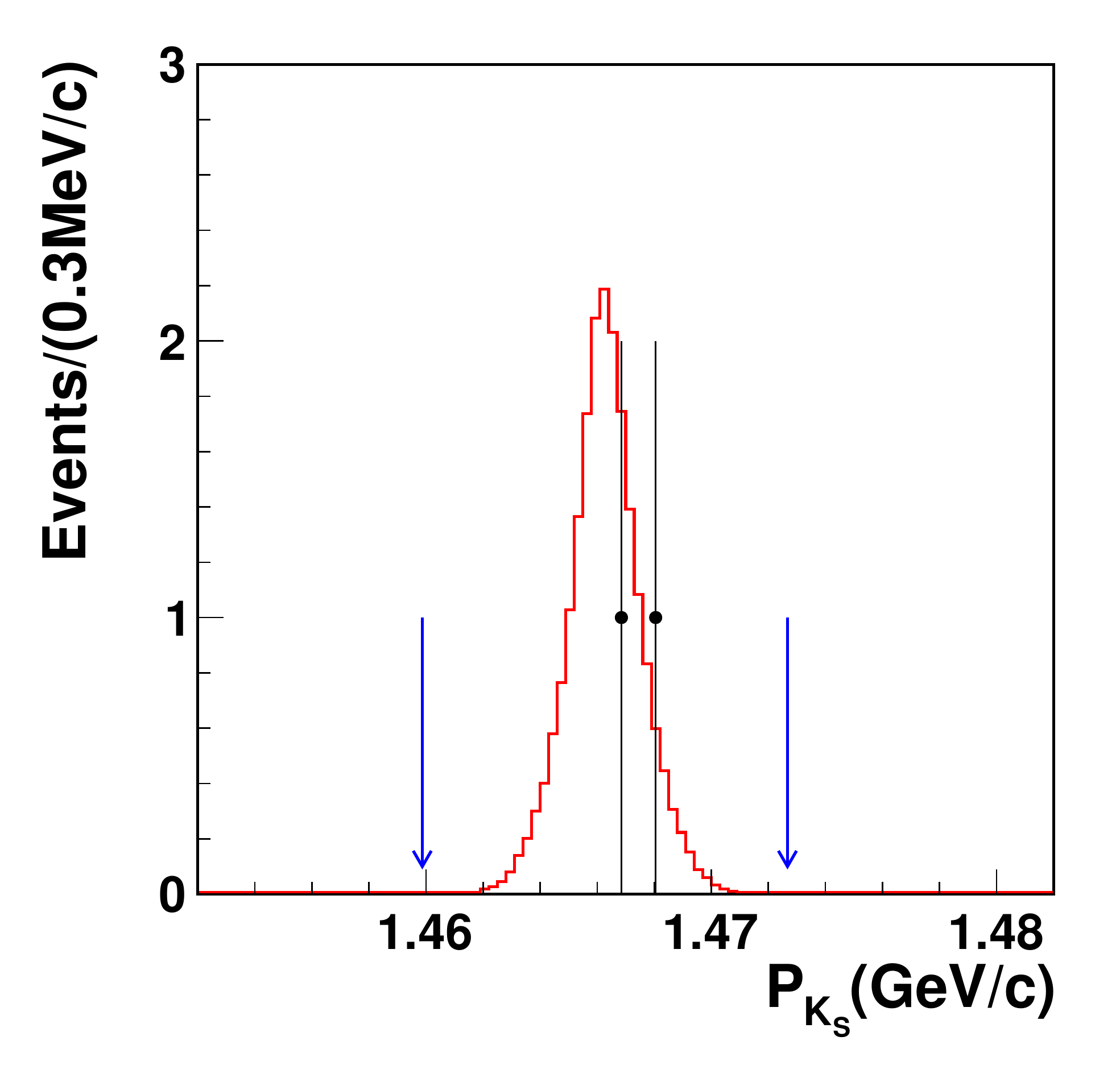}
\caption{The distribution of $\ks$ momentum in the
  $\jpsi$ rest frame.  The (black) crosses are from data,
  and the (red) solid line is from the signal MC sample. The arrows
  indicate the 5$\times\sigma_{w}$ selection region.}
\label{ksksmomdata}
\end{figure}

The $\ks\ks$ candidates are then combined in a 4C kinematic fit, where
the constraints are provided by energy and momentum conservation. Only
events with $\chi^{2}<40$ are retained. The distribution of the $\ks$
momentum in the $\jpsi$ rest frame is shown in Fig.~\ref{ksksmomdata}.
The $\ks$ momentum resolution is determined from the signal MC sample
as $\sigma_{w}=$ 1.3 MeV/$c$, which is the weighted average of the
standard deviations of two Gaussians with common mean. The number of
signal events is obtained by counting the remaining events within
$5\times\sigma_{w}$ of the expected momentum.  After all requirements
have been imposed,  two events remain in this region.

The same selection criteria are applied to the inclusive MC sample,
which shows that the background mainly comes from the processes
$\jpsi\ar\pip\pim\pip\pim$ and $\jpsi\ar\ks\kl$. Their contributions
are estimated from the corresponding MC samples using
\begin{spacing}{0.5}
  \begin{equation}~\label{bkgequ}
    N_{\rm{exp}}^{X}=N_{\jpsi}\cdot \BR(\jpsi\ar X)\cdot \epsilon^{X}_{\ks\ks},
  \end{equation}\\
\end{spacing}
\noindent
where $X$ represents the corresponding channels
$\jpsi\ar\pip\pim\pip\pim$ or $\jpsi\ar\ks\kl$($\ks\ar\pip\pim$), and
$N_{exp}^{X}$ is the expected number of events from channel
$X$. $\BR(\jpsi\ar X)$ is the product branching fractions of the
cascade decay, where $\BR(\jpsi\ar\pip\pim\pip\pim)$ is taken from the
PDG~\cite{PDG2016}, $\BR(\jpsi\ar\ks\kl)$ is set to the value obtained
in this paper, and $\epsilon^{X}_{\ks\ks}$ is the $\ks\ks$ selection
efficiency for a sample of $X$ events. The efficiencies of
$\jpsi\ar\pip\pim\pip\pim$ and $\ks\kl$ channels are
$(1.9\pm0.6)\times 10^{-7}$ and $(8.5\pm3.4)\times 10^{-6}$,
respectively.  The expected background numbers are calculated to be
$N_{\rm{exp}}^{\pip\pim\pip\pim}=0.9\pm0.3$ and
$N_{\rm{exp}}^{\ks\kl}=1.5\pm0.6$, where the uncertainties are from
propagation of the items in Eq.(\ref{bkgequ}).  Some other
exclusive processes, such as $\jpsi\ar\gamma\ks\ks$, are also studied
with high statistics MC samples, but none of them survive the event
selection.

Table~\ref{kskssys} summarizes the systematic uncertainties in the
search for $\jpsi\ar\ks\ks$. Common uncertainties including those from
the number of $\jpsi$ decays and the $\ks\ar\pip\pim$ branching
fraction are the same as described in Sec.~\ref{kskl}. The
uncertainty from $\ks$ reconstruction is evaluated according to the
$\ks$ selection criteria used in this channel, with a method similar
to that in Sec.~\ref{kskl}, and is determined to be 1.5\% per
$\ks$. The uncertainty from the 4C kinematic fit is investigated using
the control sample of $\jpsi\ar\g\ks\ks$, and the difference of the
efficiency between the data and MC samples is taken as the systematic
uncertainty associated with the kinematic fit.
\begin{table}[htpb]
    \centering
    \caption{The systematic uncertainties related to the search for
      $\jpsi\ar\ks\ks$.}
        \begin{tabular}{lc}
        \hline
        Source                      &Uncertainty~(\%)    \\
        \hline
        $\ks$ reconstruction        &3.0                \\
        4C kinematic fit            &1.1                \\
        $\BR(\ks\ar\pip\pim)$       &0.2                \\
        $N_{\jpsi}$                 &0.6                \\
        \hline
        \ST Total                   &3.2                \\
        \hline
        \\
        \end{tabular}
    \label{kskssys}
\end{table}

Since we have not observed a significant signal, an upper limit for
$\BR(\jpsi\ar\ks\ks)$ is set at the 95\% C.L.  The upper limit is
calculated using the relation
\begin{spacing}{0.5}
  \begin{equation}
    \begin{split}
      \BR(\jpsi\ar\ks\ks)&<\frac{N^{\rm{UL}}}{\epsilon_{\rm{MC}}\cdot N_{\jpsi}}.
    \end{split}
  \end{equation}\\
\end{spacing}
\noindent
where $N^{\rm{UL}}$ is the upper limit on the number of signal events
estimated with $N_{\rm{obs}}$ and $N_{\rm{bkg}}$ using a frequentist
approach with the profile likelihood method, as implemented in the ROOT
framework~\cite{trolke}, and $\epsilon_{\rm{MC}}$ is the detection
efficiency. The calculation includes statistical fluctuations and
systematic uncertainties. The signal and background fluctuations are
assumed to follow Poisson distributions, while the systematic
uncertainty is taken to be a Gaussian distribution.  The branching
fraction of $\ks\ar\pip\pim$ is included in the event selection
efficiency $\epsilon_{\rm{MC}}$.  The values of variables used to
calculate the upper limit on the branching fraction and the final
result are summarized in Table~\ref{ksksresult}, where the
$N_{\rm{bkg}}$ is the sum of $N_{\rm{exp}}^{\pip\pim\pip\pim}$ and
$N_{\rm{exp}}^{\ks\kl}$.

\begin{table}[htpb]
    \centering
    \caption{Numbers used in the $\BR(\jpsi\ar\ks\ks)$ calculation of the upper limit on the signal yield at the 95\% C.L.}
        \begin{tabular}{lc}
        \hline
        $N_{\rm{obs}}$  \STi                      &2                                   \\
        $N_{\rm{bkg}}$  \STi                      &2.4                                 \\
        $N^{\rm{}UL}$   \STi                      &4.7                                 \\
        $\epsilon_{\rm{MC}}$(\%)  \STi            &25.7                                \\
        \hline
        $\BR(\jpsi\ar\ks\ks)$~(95\% C.L.)  \STi  &$<1.4\times 10^{-8}$                \\
        \hline
        \end{tabular}
    \label{ksksresult}
\end{table}

\section{Summary}\label{Summary}

Based on a data sample of $1.31\times 10^{9}$ $\jpsi$ events collected
with the BESIII detector, the measurements of $\jpsi\ar \ks\kl$ and
$\ks\ks$ have been performed. The branching fraction of
$\jpsi\ar\ks\kl$ is determined to be
$\BR(\jpsi\ar\ks\kl)=(1.93\pm
0.01~(\rm{stat.})\pm0.05~(\rm{syst.}))\times 10^{-4}$, which agrees
with the BESII measurement~\cite{ksklbes2} while discrepancy with the
CLEO data~\cite{kkcleo} persists.  Compared with the world average
value listed in the PDG~\cite{PDG2016}, the relative precision is
greatly improved, while the central value is consistent.  With regard
to the search for the \CP and Bose-Einstein statistics violating
process $\jpsi\ar\ks\ks$, an upper limit on its branching fraction is
set at the 95\% C.L. to be $\BR(\jpsi\ar\ks\ks)<1.4\times 10^{-8}$,
which is an improvement by 2 orders in magnitude compared to the
best previous searches~\cite{kkmark,ksksbes2}.  The upper limit
reaches the order of the EPR expectations\cite{roos}.

\section{Acknowledgment}
\label{sec:Acknowledgement}

The BESIII collaboration thanks the staff of BEPCII and the IHEP
computing center for their strong support. This work is supported in
part by National Key Basic Research Program of China under Contract
No. 2015CB856700; National Natural Science Foundation of China (NSFC)
under Contracts No. 11235011, No. 11335008, No. 11425524,
No. 11625523, No. 11635010; the Chinese Academy of Sciences (CAS)
Large-Scale Scientific Facility Program; the CAS Center for Excellence
in Particle Physics (CCEPP); Joint Large-Scale Scientific Facility
Funds of the NSFC and CAS under Contracts No. U1232105, No. U1332201,
No. U1532257, No. U1532258; CAS under Contracts No. KJCX2-YW-N29,
No. KJCX2-YW-N45, No. QYZDJ-SSW-SLH003; 100 Talents Program of CAS;
National 1000 Talents Program of China; INPAC and Shanghai Key
Laboratory for Particle Physics and Cosmology; German Research
Foundation DFG under Contracts No. Collaborative Research Center CRC
1044, No. FOR 2359; Istituto Nazionale di Fisica Nucleare, Italy;
Joint Large-Scale Scientific Facility Funds of the NSFC and CAS;
Koninklijke Nederlandse Akademie van Wetenschappen (KNAW) under
Contract No. 530-4CDP03; Ministry of Development of Turkey under
Contract No. DPT2006K-120470; National Natural Science Foundation of
China (NSFC) under Contract No. 11505010; National Science and
Technology fund; The Swedish Resarch Council; U. S. Department of
Energy under Contracts No. DE-FG02-05ER41374, No. DE-SC-0010118,
No. DE-SC-0010504, No. DE-SC-0012069; University of Groningen (RuG)
and the Helmholtzzentrum f\"ur Schwerionenforschung GmbH (GSI),
Darmstadt; and WCU Program of National Research Foundation of Korea
under Contract No. R32-2008-000-10155-0.

\end{document}